The Airy fibre: an optical fibre that guides light diffracted by a circular aperture

I. Gris-Sánchez,[1] D. Van Ras,[2] T. A. Birks[1,*]

[1]Department of Physics, University of Bath, Claverton Down, Bath BA2 7AY, United Kingdom
[2]Specialty Department, Draka Comteq Fiber BV, Zwaanstraat 1 5651 CA Eindhoven, Netherlands

*Corresponding author: t.a.birks@bath.ac.uk

We have designed and made an optical fibre that guides an approximate Airy pattern as one of its guided modes. The fibre's attenuation was 11.0 dB/km at 1550 nm wavelength, the match between the fibre's mode and the ideal infinite Airy pattern was 93.7%, and the far field resembled a top-hat beam. The guidance mechanism has strong similarities to photonic bandgap guidance.

## 1. INTRODUCTION

When a plane wave of light is obstructed by a circular aperture, the transmitted light evolves in the far field into a characteristic pattern with a bright central "Airy disc" surrounded by a set of ever-fainter concentric rings [1,2]. This is known as the Airy pattern, Fig. 1.

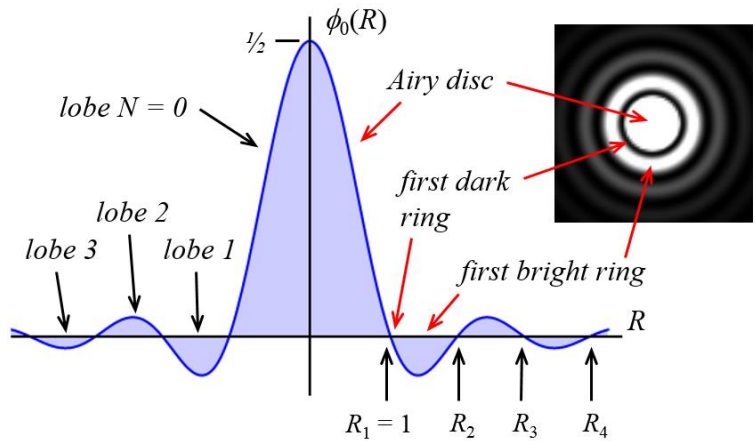

Fig. 1. The field $\phi_0(R)$ of the ideal Airy pattern, with a 2-D intensity pattern (centre saturated) inset. The radial coordinate R is normalised so that the first dark ring lies at $R_1 = 1$. The Airy disc is lobe $N = 0$, the first bright ring is lobe $N = 1$, etc. Here and elsewhere we include negative values of R for visual effect; the absolute value of R is taken.

For example, it appears in the images of stars formed by telescopes (in the absence of aberrations caused by the instrument or the atmosphere). This light can be transported to remote instruments such as spectrographs using an optical fibre. A multimode fibre can capture essentially all of the light, but at the cost of degrading its focal ratio or etendue. To conserve etendue and mode purity, a single-mode fibre (SMF) can be used. However, the guided mode of SMFs differs from the Airy pattern, such that at most about 80% of the light can be coupled into the fibre even in theory [3,4]. This can be a serious handicap in light-starved applications such as astronomy, optical resonators (including lasers) and single-photon quantum optics experiments incorporating an aperture.

Here we report the design and fabrication of an optical fibre that guides a close approximation to the Airy pattern as one of its modes. The mode of the fabricated fibre extends to the third ring of the pattern, which it matches closely enough to capture 94% of the light. The design procedure is an example of a general method for designing optical fibres (and indeed other waveguides) that support a good approximation to any given field distribution as a guided mode [5].

## 2. FIBRE DESIGN

### A. The inverse waveguide problem and the Airy pattern

In a typical "forward" waveguide problem, the refractive index distribution $n(r,\theta)$ is known and the field distributions $\psi(r,\theta)$ and propagation constants $\beta$ of the guided modes are found from the scalar wave equation

$$\nabla_T^2 \psi + (k^2 n^2 - \beta^2)\psi = 0 \quad (1)$$

where $\nabla_T^2$ is the transverse Laplacian operator, $k = 2\pi/\lambda$ and $\lambda$ is the wavelength. The weak-guidance approximation (valid for the small index variations typical of silica fibres) allows the vector direction of the field to be eliminated from the problem [6-9]. The differential equation must be solved to find $\psi(r,\theta)$ and $\beta$, with $\psi$ and its gradient continuous at any boundaries.

The inverse problem of finding a waveguide $n(r,\theta)$ that guides a desired mode $\psi(r,\theta)$ is in principle much simpler, since Eq. (1) can be rearranged to give

$$n^2(r,\theta) = \frac{-\nabla_T^2 \psi(r,\theta)}{k^2 \psi(r,\theta)} + n_{eff}^2 \quad (2)$$

where the mode's effective index $n_{eff} = \beta/k$ can be treated as an arbitrary offset constant. Unlike the forward problem there is no differential equation to be solved; we merely substitute the known $\psi(r,\theta)$.

Eq. (2) can be normalised by a length scale $a$, such as a core radius. In the case of a circularly-symmetric ($l = 0$) mode $\psi(R)$ of a circular fibre, where there is no $\theta$ dependence, this yields

$$v(R) = \frac{-\left\{\frac{d^2\psi}{dR^2} + \frac{1}{R}\frac{d\psi}{dR}\right\}}{\psi} \quad (3)$$

where

$$v(R) = a^2 k^2 (n^2 - n_{eff}^2) \quad (4)$$

is a normalised index distribution and $R = r/a$ is a normalised radial coordinate. (In a uniform region, $v$ matches the waveguide parameter $U^2$ [6].)

We want to apply this method to the Airy pattern, which is the Fourier transform of a 2-D circular top-hat aperture function [1,2]. It is circularly symmetric and has the form

$$\phi_0(R) = \frac{J_1(j_{11}R)}{j_{11}R} \quad (5)$$

where $J_l$ is the Bessel function of the first kind of order $l$ and $j_{lm}$ is its $m$-th zero. As $R \to 0$, $\phi_0(R) \to \tfrac{1}{2}$. Its absolute scale will depend on the optics and the wavelength and can be adjusted using lenses, but in Eq. (5) we have defined $a$ to be the absolute radius of the first dark ring at $R = R_1 = 1$. The pattern's zeros at $R = R_m = j_{1m}/j_{11}$ bound its bright lobes: lobe $N = 0$ is the central Airy disc within $R_1 = 1$, and lobe $N \geq 1$ is the $N$-th bright ring bounded by $R_N$ and $R_{N+1}$, Fig. 1. Unfortunately, the substitution $\psi(R) = \phi_0(R)$ in Eq. (3) produces infinite values of refractive index at the zeros because of the denominator $\psi$, Fig. 2(a).

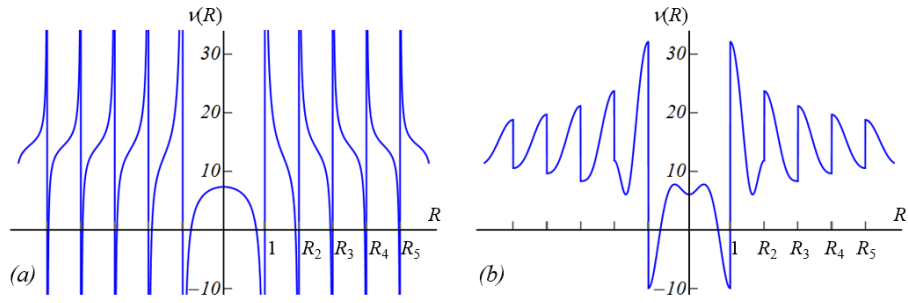

Fig. 2. The normalised index distribution $v(R)$ calculated using Eq. (3) for (a) the exact Airy pattern of Eq. (5) and (b) the Airy mode of Eq. (8) and Table 1. The refractive index in the "Airy cladding" $R > 1$ is notably high compared to that in the Airy disc $R < 1$.

## B. Approximation 1: eliminating infinite refractive indices

We avoided the problem by finding a close approximation $\psi(R)$ to the Airy pattern (the "Airy mode") that does not require infinities in refractive index. There are many ways to do this. Our approach was to define separate approximations $\psi_N(R)$ in each lobe $N$, and require them to satisfy the following constraints:

I. The numerator in Eq. (3) is zero wherever the field is zero. This permits $v(R)$ to be finite.
II. The field is zero at the zeros of the ideal Airy pattern $\phi_0(R)$.
III. The field is finite and its slope is zero at the origin.
IV. The slope of the field matches that of the Airy pattern at its zeros.
V. The field matches that of the Airy pattern at the origin.

As well as helping to make $\psi_N(R)$ look like $\phi_0(R)$, constraints II-V ensure that the boundary conditions required of any scalar waveguide mode are automatically satisfied: that the field and its slope are finite and continuous at every boundary [6,8]. Furthermore, matching the fields at a boundary to $\phi_0(R)$ rather than directly to each other makes each lobe's solution independent of the others, which is a great simplification.

Extra constraints can be imposed to make the mode a better match to $\phi_0(R)$. We added just one more:

VI. The field matches the Airy pattern at one additional point in each lobe. In most cases this was the midpoint between the lobe's bounding zeros $R_N$ and $R_{N+1}$ (or between 0 and $R_1$ for $N=0$). However, for $N=1$ the location of the peak of $|\phi_0(R)|$, at $R = 1.3403$, provided a better fit.

The simplest way to satisfy constraint I is to express $\psi_N(R)$ using basis functions $f(R)$ satisfying

$$\frac{d^2 f}{dR^2} + \frac{1}{R}\frac{df}{dR} + P^2 f = 0 \qquad (6)$$

Eq. (6) is Bessel's equation of order zero, with a general solution proportional to

$$f(R) = J_0(PR) + CY_0(PR) \qquad (7)$$

where $Y_0$ is the Bessel function of the second kind of order 0. There is a continuum of such functions $f(R)$ for all values of the constants $C$ and $P$. However, constraints II and III act as boundary conditions on Eq. (7) and restrict $P$ to a discrete set of values $P_{N,i}$ (where $N$ is the lobe number and $i$ indicates the $i$-th smallest positive solution for $P$), with corresponding values $C_{N,i}$ for $C$. Numerical solutions are given in Table 1 for the values of $N$ and $i$ relevant to our experiments. (Numerical methods were not needed for $N=0$, where $C_{0,i} = 0$ and $P_{0,i} = j_{0i}$ directly.)

*Table 1. Constants P, C and A in Eqs. (7) and (8) in each lobe N = 0 to 3, for our M = 3 approximation to the Airy pattern*

| N | i | $P_{N,i}$ | $C_{N,i}$ | $A_{N,i}$ |
|---|---|-----------|-----------|-----------|
| 0 | 1 | 2.4048 | 0 | 0.45734 |
|   | 2 | 5.5201 | 0 | 0.063497 |
|   | 3 | 8.6537 | 0 | −0.020836 |
| 1 | 1 | 3.7637 | 5.0501 | 0.035422 |
|   | 2 | 7.5527 | −1.9797 | 0.018511 |
|   | 3 | 11.336 | −0.48985 | −0.0097653 |
| 2 | 1 | 3.8053 | 8.3602 | 0.013937 |
|   | 2 | 7.6205 | −1.4872 | −0.0086658 |
|   | 3 | 11.434 | −0.28498 | −0.0016756 |
| 3 | 1 | 3.8177 | 11.577 | 0.0073579 |
|   | 2 | 7.6406 | −1.3281 | 0.0049240 |
|   | 3 | 11.462 | −0.20373 | −0.00065756 |

Writing the Airy mode $\psi_N(R)$ in lobe N as a sum of the first M of that lobe's basis functions $f_{N,i}(R)$

$$\psi_N(R) = \sum_{i=1}^{M} A_{N,i} f_{N,i}(R) \qquad (8)$$

with $P_{N,i}$ and $C_{N,i}$ from Table 1 implicit in $f_{N,i}(R)$, meets constraints I, II and III. This is like a truncated Fourier series, with the $f_{N,i}(R)$ being the Bessel-function equivalents of sinusoidal harmonics between zeros at $R_N$ and $R_{N+1}$, Fig. 3(a). M further boundary conditions (beyond constraints I, II and III) yield M simultaneous linear equations from Eq. (8), to solve for the M unknown "Fourier coefficients" $A_{N,i}$ for each lobe. Taking $M \to \infty$ terms in the series recovers the exact Airy pattern $\phi_0(R)$ but also the infinities in Eq. (3) that we were trying to get rid of in the first place. This is why we adopted the fewest further constraints (IV, V and VI) to provide a good-enough match to $\phi_0(R)$. These impose three conditions on the field in each lobe (so $M = 3$), giving the $A_{N,i}$ values in Table 1.

The Airy mode field $\psi(R)$ of Eq. (8) can now be substituted into Eq. (3) to give the normalised index profile $v(R)$ of a fibre that guides it, Fig. 2(b). The profile, though complicated, is everywhere finite and so is physically realisable. The central part (the Airy disc) is depressed in index compared to its surroundings (the "Airy cladding"). Indeed, the mode's field is not evanescent anywhere in the Airy cladding, where $v(R) > 0$, since the horizontal axis $v = 0$ marks the effective index of the mode ($n = n_{eff}$ in Eq. (4)). Our Airy mode is therefore guided by a mechanism akin to a photonic bandgap, like an aperiodic graded-index Bragg fibre [10-13].

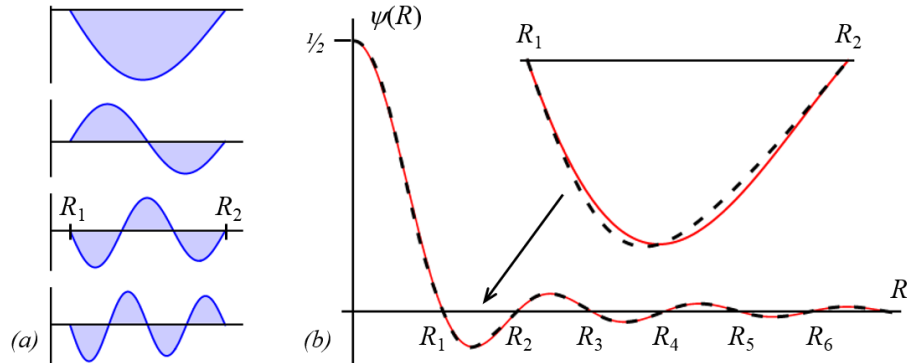

Fig. 3. (a) The first 4 basis functions $f_{N,i}(R)$ in lobe $N = 1$. (b) Field distributions of the ideal Airy pattern (solid red curve) and our Airy mode (dashed black curve). The inset magnifies lobe 1.

Fig. 3(b) compares the Airy mode $\psi(R)$ of Eq. (8) to the ideal Airy pattern $\phi_0(R)$ of Eq. (5). They are difficult to distinguish: a small change in the field corresponds to a large change in the index profile, Fig. 2. The match between the fields is quantified by an overlap integral

$$\Omega = \frac{\left\{\int \Psi_1 \Psi_2 \, dA\right\}^2}{\left\{\int \Psi_1^2 \, dA\right\}\left\{\int \Psi_2^2 \, dA\right\}} \quad (9)$$

representing the fraction of the power in mode $\Psi_2$ excited by mode $\Psi_1$ in an idealised butt-coupling experiment [8,9]. For $\psi(R)$ and $\phi_0(R)$, $\Omega = 99.92\%$; an excellent match.

## C. Approximation 2: terminating the profile in the radial direction

The Airy pattern extends to infinite $R$ but real fibres are finite. Therefore we must truncate the index profile at a suitable $R = R_c$, beyond which we impose a uniform cladding $v(R) = v_c$. The mode will not match the Airy pattern here, but if $R_c$ is big enough the difference is unimportant. However, $v_c$ should be negative so that the field is evanescent in the uniform cladding and not leaky ($n < n_{eff}$ in Eq. (4)). In that case the field here is

$$\psi_c(R) = D K_0(WR) \quad (10)$$

like any circularly-symmetric evanescent field in a uniform cladding [6-8]. $K_l$ is the modified Bessel function of the second kind of order $l$, $W = (-v_c)^{1/2}$ is the familiar cladding parameter of waveguide theory [6] and $D$ is an amplitude constant.

We are not free to truncate the profile arbitrarily [5]. Matching the fields $\psi_c(R)$ and $\psi(R)$ and their slopes at $R = R_c$ yields

$$\frac{1}{\psi(R_c)} \left.\frac{d\psi}{dR}\right|_{R=R_c} = -W \frac{K_1(WR_c)}{K_0(WR_c)} \quad (11)$$

an implicit relationship between $R_c$ and $v_c$ plotted in Fig. 4(a) for lobe 3. Valid solutions exist only for the outer part of each lobe, where the field declines towards zero from its extremum (otherwise it matches a growing rather than a decaying field in the cladding). Across this range, $v_c$ decreases from 0 (putting the Airy mode at cutoff) to $-\infty$ (the "infinite well" limit). We can therefore choose any desired negative $v_c$ but must truncate at the right corresponding $R_c$, a unique location in each lobe.

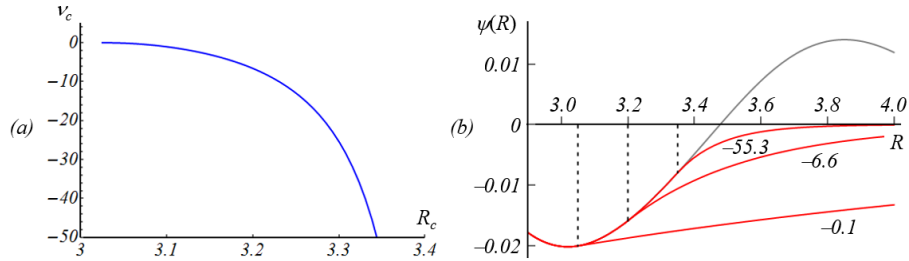

Fig. 4. (a) Normalised index $v_c$ of the uniform cladding versus truncation radius $R_c$, for the $N = 3$ lobe. (b) Field distributions $\psi(R)$ for Airy modes (red curves) truncated at three different $R_c$ (dashed lines), together with the untruncated Airy mode (grey curve). The curves are labelled with the corresponding $v_c$, which can be compared to the vertical scale of Fig. 2(b).

The effect of this choice is shown in Fig. 4(b) for the $N = 3$ example. A more negative $v_c$ shrinks the evanescent field and pushes $R_c$ outwards. This improves the fit to $\phi_0(R)$, since the evanescent field has opposite sign to $\phi_0(R)$ in the next lobe and so reduces $\Omega$. However, a more negative $v_c$ increases the number of other modes supported by the fibre. It also increases the overall range of refractive index, making fibre fabrication more challenging or even impossible in practice. We therefore chose to truncate $\psi(R)$ in the $N = 3$ lobe at $R_c = 3.2$, the middle case in Fig. 4(b). This gives $v_c = -6.6$, which close to (but greater than) the minimum value of $v(R)$ in the rest of the profile and so does not increase the index range across the fibre, Fig. 2(b). Table 2 summarises the key parameters for this choice, with $D$ found from Eq. (10) at $R = R_c$ (where $\psi_c$ matches $\psi$ from Eq. (8)).

Table 2. Parameters defining our truncation of the Airy mode by a uniform cladding, using Eqs. (10) and (11) and the untruncated field of Eq. (8) and Table 1

| $R_c$ | $W$ | $v_c$ | $D$ |
|---|---|---|---|
| 3.2 | 2.5694 | −6.6020 | −137.22 |

Our final truncated Airy mode $\psi(R)$ is plotted in Fig. 5(a). Fig. 5(b) is the corresponding index profile $v(R)$, differing from Fig. 2(b) by the imposition of the uniform cladding. The overlap between the mode and the ideal Airy pattern is $\Omega = 94.64\%$, meaning that the mode could capture up to this fraction of the power in the Airy pattern. For comparison, 95.29% of the power in an Airy pattern resides in lobes 0 to 3, so nearly all the light not captured by our mode is outside the third ring anyway. The overlap can always be improved by truncating the index profile further out, but at the cost of a bigger fibre.

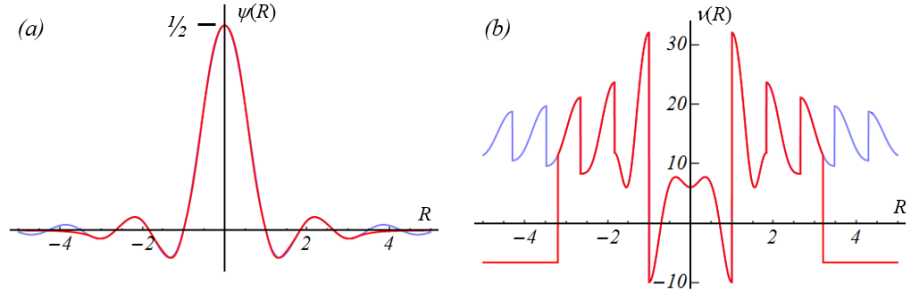

Fig. 5. (a) Field distributions of the truncated Airy mode (red) and the ideal Airy pattern it approximates (blue). The middle curve of Fig. 4(b) enlarges the truncation region. (b) Normalised index distributions of fibres guiding the truncated (red) and untruncated (blue) Airy modes.

### D. Other modes and wavelengths

The Airy mode is not the fibre's fundamental mode: it has zeros in its field distribution, whereas a fundamental mode does not [8,9]. Indeed there is plenty of high-index material in the Airy cladding to support lower-order modes. The truncated Airy mode of Fig. 5 has three zeros (the three remaining dark rings) and so must be the 4th $l = 0$ circularly-symmetric mode: the $LP_{04}$ mode [8,9]. Given our index profile, Eq. (1) was solved using conventional methods to find a total of 55 spatial modes, of which ours is the 40th mode in decreasing order of $\beta$. However, all the lower-order modes are concentrated in the high-index rings beyond $R = 1$; the Airy mode is the first to be concentrated in the low-index material within $R = 1$. Some of them are shown in Fig. 6.

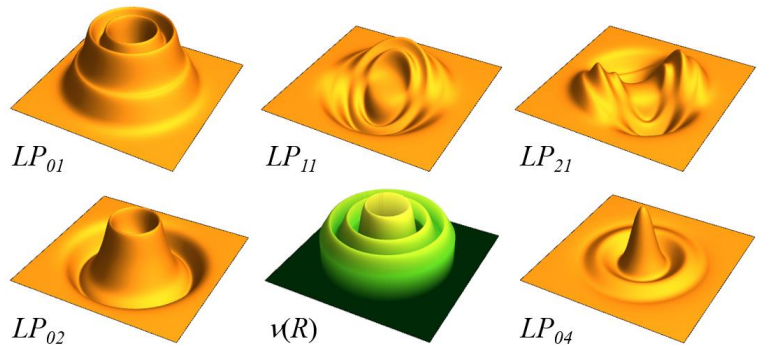

Fig. 6. Field distributions of selected $LP_{lm}$ modes of the truncated Airy fibre, together with $v(R)$ of Fig. 5(b) for scale. The fundamental mode ($LP_{01}$) has no zeros and the Airy mode ($LP_{04}$) has three annular zeros.

We can also solve Eq. (1) to see how the mode changes with wavelength. The presence of $k$ in Eq. (2) means that a given index distribution $n(R)$ can only guide our specified $\psi(R)$ at a single design wavelength $\lambda_0$. Nevertheless we found that the mode remains a good approximation to the ideal Airy pattern $\phi_0(R)$ for a significant range of $\lambda$, Fig. 7(a). A dispersion plot of all $l = 0$ modes, Fig. 7(b), shows that the curve for the $LP_{04}$ Airy mode has a shallower slope, characteristic of light in the low-index central region rather than the high-index rings. There is however an anti-crossing event [14] at $\lambda_0/\lambda \approx 1.6$ where the most "Airy-like" mode changes identity to $LP_{05}$. (There are others at higher frequencies.) Such anti-crossings are familiar in the dispersion plots of photonic bandgap fibres [15,16]. The high-$\Omega$ range in Fig. 7(a) is bounded at short wavelengths by the anti-crossing and at long wavelength by the mode's impending cutoff.

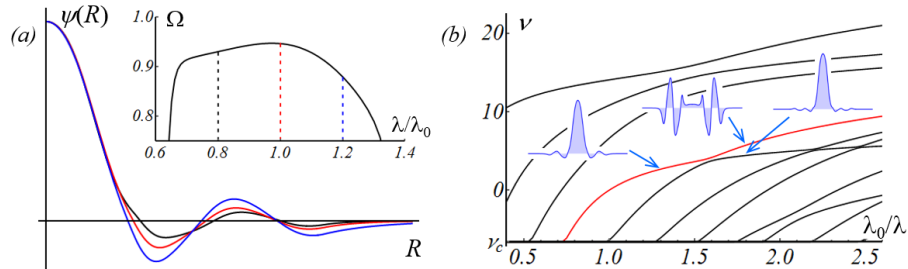

Fig. 7. (a) The Airy mode at the design wavelength $\lambda_0$ (red), $0.8\lambda_0$ (black) and $1.2\lambda_0$ (blue). Inset: the overlap $\Omega$ with the ideal Airy pattern $\phi_0(R)$ versus wavelength $\lambda/\lambda_0$. The dashed lines mark the wavelengths in the main plot. (b) The normalised effective index $v$ of the fibre's $l = 0$ modes versus frequency $\lambda_0/\lambda$. The red curve is the Airy mode ($LP_{04}$). Insets: mode fields either side of the first anti-crossing, where the most Airy-like mode changes identity.

## 3. EXPERIMENTAL FIBRE

### A. Fibre fabrication

To make a real Airy fibre, the normalised index profile was converted into a physical fibre design using Eq. (4), given suitable choices for $a$, $k = 2\pi/\lambda_0$ and the unnormalised refractive index $n_0$ corresponding to a specific normalised value $v_0$

$$n^2(r) = n_0^2 + \frac{v(r/a) - v_0}{a^2 k^2} \qquad (12)$$

We made one minor change to the design plotted in Fig. 5(a) and summarised in the tables. By adjusting the point where criterion VI was applied in lobe $N = 1$ to $R = 1.5052$ we reduced the required index variation across the fibre by 12.5%, at the cost of a reduction in $\Omega$ of just $2\times10^{-5}$. The only affected design parameters were $(A_{11}, A_{12}, A_{13}) = (0.037719, 0.018506, -0.0036321)$ in Table 1. The index profile for $2a = 13.5$ μm, $\lambda_0 = 1550$ nm and a maximum index $n_0 = 1.44$ (undoped fused silica) is plotted as the red curve in Fig. 8(a). The overall index variation corresponds to a numerical aperture of NA = 0.22, which is realistic for silica fibre fabrication technology.

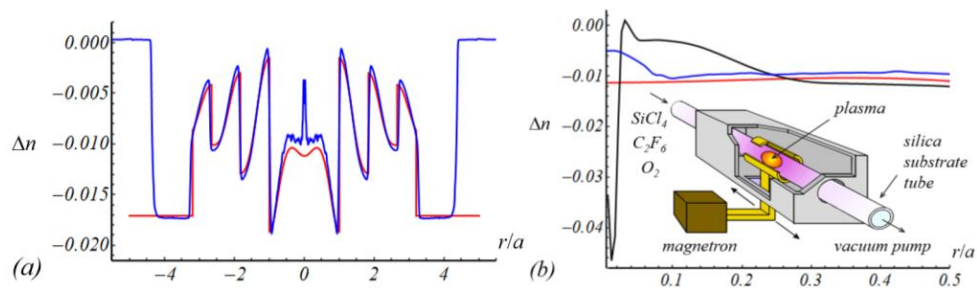

Fig. 8. (a) Physical refractive index profiles, relative to undoped silica, of (red) the final Airy fibre design and (blue) the measured PCVD preform. The radial coordinate $r$ is normalised to $a$, the radius of the Airy disc, and negative $r$ is a copy of positive $r$ for visual effect. (b) As (a) for just the centre of the preform, including a trial preform (black) made without burn-off etching: contrast the vertical scale with (a). Inset: a schematic diagram of the PCVD process.

We made the preform for our fibre by plasma-activated chemical vapour deposition (PCVD) [7,17], Fig 8 (inset), which is particularly suitable for fabricating fibres with complex refractive index profiles such as ours. Continuously-controlled proportions of vapour-phase reactants (SiCl$_4$, GeCl$_4$, C$_2$F$_6$ and O$_2$) are passed at low pressure into a silica substrate tube heated to roughly 1200 °C, where they are ionised in a microwave plasma. By repeatedly scanning the plasma along the tube, thin layers of doped silica glass are directly formed on the inside. The refractive index of each layer is determined by the concentration of the Ge (germanium) and F (fluorine) dopants, which respectively increase and decrease the index. In this way, an intricate graded-index profile can be accurately built up from many thousands of thin layers, each with its own composition and hence refractive index.

In our case, we chose to vary the index using F-doping alone. This is because it is easier to control the deposition of F than Ge, especially at sharp steps in the index profile like those in our fibre design. The uniform deposition rate of F-doped silica allowed a fully-automated mass flow control scheme to be used. Because F can only decrease the index compared to undoped silica, we set the maximum index in our profile to match undoped silica. The uniform cladding region of our fibre beyond $R = R_c$ therefore needed to be a low-index (F-doped) "trench" between the graded-index (also F-doped) Airy cladding layers and a high-index (undoped) outer region derived from the preform's pure-silica substrate tube.

The total PCVD process time was 6 hours, after which the tube was collapsed using an electric furnace to produce a solid preform suitable for drawing to fibre. The raised temperature can cause some of the dopants on the inside surface of the tube to evaporate. The dopants can then be re-deposited further along the tube. The results are irregularities of refractive index in the centre of the preform. This "burn off" contamination was removed by passing fluorinated gases along the tube as it collapses, to etch the inner layers [18]. Fig. 8(b) shows how the etching process greatly reduced the unwanted positive and negative excursions of refractive index due to burn off.
The refractive index distribution of the final preform is plotted in Fig. 8 along with the design profile. The match is excellent, given the complexity of the design profile and the fabrication process. 190 m of Airy fibre with an outer diameter of 97.5 µm was then drawn from this preform in the conventional way [7,9].

## B. Fibre characterisation

Assuming the whole index profile of the preform scales in proportion and comparing with the design profile of Fig. 8, the measured outer diameter implies a core diameter of $2a = 15.6$ µm and an estimated optimal operating wavelength of $\lambda_0 = 1790$ nm - somewhat greater than intended. An optical micrograph of the fibre cross-section is shown in Fig. 9, along with a cross-sectional intensity plot from the image. Waveguiding concentrates the shorter-wavelength visible light in the higher-index regions of an optical fibre illuminated in transmission, and it is notable how well the features of the index profile of Fig. 8 are replicated in the micrograph.

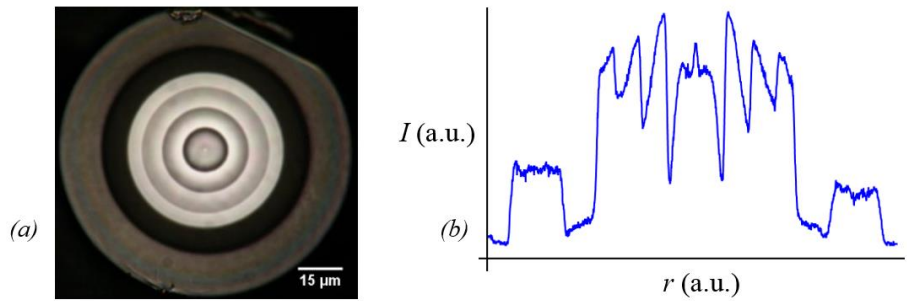

*Fig. 9. (a) Optical micrograph of one end of ~3 cm of the Airy fibre, illuminated by the microscope's white light source at the other end. (b) A cross-sectional intensity plot derived from (a) along a row of pixels through the centre.*

Light of 1550 nm wavelength was launched into a few metres of single-mode fibre (Corning SMF28e), the output of which was butt-coupled to the input of the Airy fibre. The guided mode of the SMF is of similar size to the Airy disc, favouring excitation of the Airy mode when properly aligned. The near-field intensity pattern at the output end of the Airy fibre was imaged onto an IR camera using a microscope objective, and the input was adjusted to produce what looked like an Airy pattern, Fig. 10(a). The insensitivity of the output pattern to gentle perturbation of the fibre or to shortening it (in stages) to less than 1 m suggests that the great majority of the light was in one mode. Because the dynamic range of the camera was limited, a satisfactory image of the faint rings in the pattern required the central bright disc to be saturated. We therefore recorded a second image with a known neutral density filter at the input to give an unsaturated image of the disc. We also took a dark-field image to establish the zero value in the recorded image files. We combined the three files to give a single map of the whole intensity distribution.

Quantifying the fit between the measured near-field pattern and the ideal Airy pattern requires knowledge of its phase distribution. Without a suitable interferometric method, we simply took the square root of our intensity map (to convert it to field) then negated the values in the odd-numbered rings (assuming they were of opposite phase to the others) to yield a measured mode field distribution $\psi$, Fig. 10(b). Its overlap with the ideal Airy pattern (optimally scaled) was calculated to be $\Omega = 93.7\%$, close to the simulated value of $\Omega = 94.6\%$ given the imperfection of any fibre fabrication process. The difference between the optimal wavelength of $\lambda_0 = 1790$ nm and the measurement wavelength of $\lambda = 1550$ nm will degrade the overlap slightly, but the dependence is gradual on the short wavelength side ($\lambda/\lambda_0 = 0.87$ in Fig. 7(a))

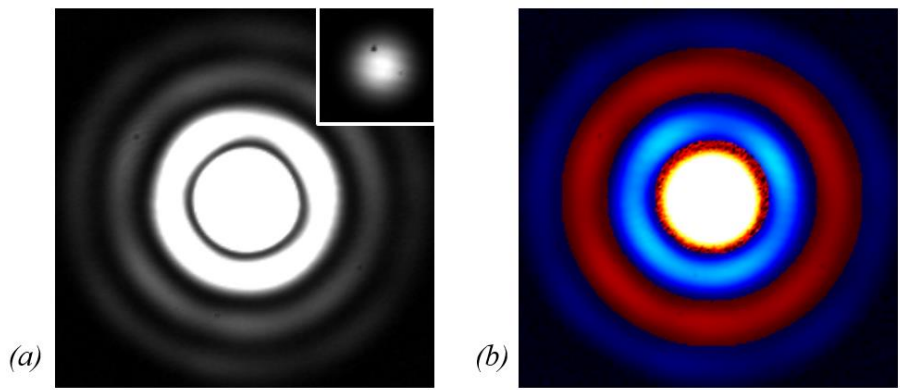

*Fig. 10. (a) Measured near-field image of the output of the fibre when trying to excite the Airy mode. The minimum intensity in the first dark ring is ~0.1% of the peak intensity - it still looks dark even though the centre of the image is saturated. Inset: an unsaturated image of the central bright disc, to the same scale. (b) The mode field $\psi$ as deduced from the measurement, assuming that the field changes sign in successive rings. Red and blue represent positive and negative phase respectively, while white means the plot is saturated.*

With the stable excitation of a single mode, we were able to measure its attenuation by the cutback method [7,8] to be 11.0 dB/km at 1550 nm wavelength, for long and short fibre lengths of 134.5 m and 34.5 m respectively. This appears to be dominated by leakage loss across the low-index "trench" of the uniform cladding to the high-index region at the outside of the fibre, Fig. 8. We used a finite-element method [19] to calculate a leakage loss of 11 dB/km for a trench width of 8.08 µm, compared with the actual trench width (deduced from the preform profile and the fibre's outer diameter) of 8.55 µm. Widening the trench would reduce the loss of the Airy mode, but the losses of other modes may decrease even more because their fields are concentrated further out than the Airy mode, Fig. 6. It is therefore questionable whether a wider trench would be beneficial.

Other modes appeared if the input excitation was misaligned or the fibre was sharply bent to cause mode coupling. We did not systematically study them, but typical images at the output when they were present are shown in Fig. 11. The other modes introduced azimuthal variations to the bright rings, and made them much more intense than they are in an Airy pattern.

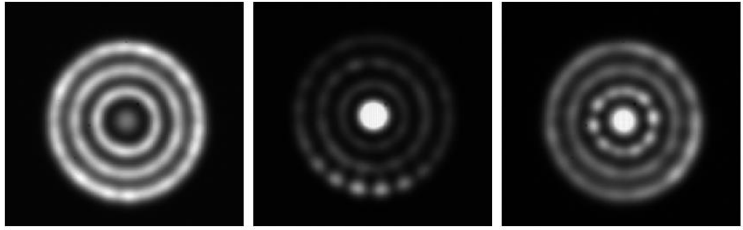

*Fig. 11. Measured near-field images when the fibre was sharply bent to couple light to other modes. These can be contrasted with the Airy mode in Fig. 10(a).*

Another test of the quality of the Airy mode is to examine the far-field output. Since the Airy pattern is the result of Fraunhofer diffraction from a circular top-hat aperture, the inverse Fourier transform relationship means that the far field from an Airy pattern should be a top-hat distribution [1,2]. The theoretical far-field intensity distributions for the Airy pattern when untruncated and truncated to three rings (calculated by squaring the Fourier transforms) are shown in Fig. 12(a) and (b) respectively. The effect of truncation is like truncating a Fourier series, and results in dips in the centre and in a ring. The far field of the Airy mode was measured by removing the microscope objective from our experiment and simply allowing the output light to diffract into free space before detection by the camera. The far-field pattern, Fig. 12(c), is actually closer to a top-hat distribution than the pattern calculated for a truncated Airy pattern, Fig. 12(b), though the central and ring-shaped dips are still present. As well as confirming the correct qualitative behaviour and validating the calculation leading to Fig. 10(b), this shows how the Airy fibre can be used to transport light and generate an approximate top-hat beam at the output far field, perhaps as an alternative to fibre designs that transport a top-hat beam in the near field [5,20,21].

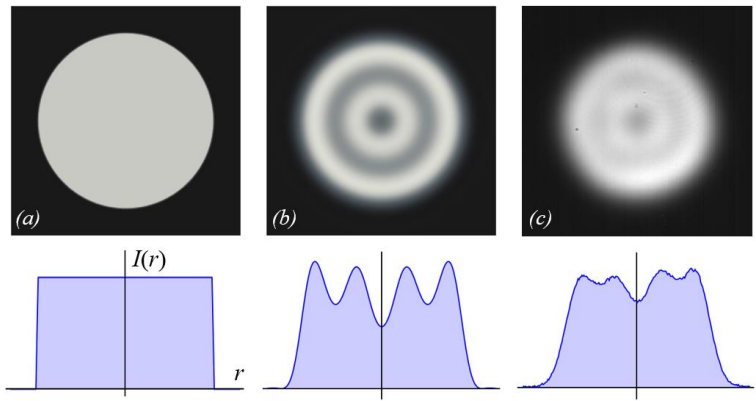

*Fig. 12. Far-field intensity distributions (a) calculated for an ideal Airy pattern, (b) calculated for an Airy pattern truncated to the first three rings, and (c) measured for the Airy mode of our fibre. The (lower) 1-D plots are horizontal slices through the centres of the (upper) 2-D plots.*

## 4. CONCLUSIONS

We have designed and fabricated an optical fibre that guides a good approximation to the Airy pattern as one of its guided modes. The experimental fibre was fabricated from a fluorine-doped silica PCVD preform and its attenuation was 11 dB/km. Limiting the design to the field within the first three bright rings limits the attainable overlap with the ideal Airy pattern to 95.3% in theory and 93.7% in practice. The fibre's far-field pattern resembled the expected top-hat beam profile.

The Airy mode is not the fundamental mode of the fibre. The central core-like region where its power is concentrated is of lower refractive index than the surrounding cladding. In an untruncated fibre the Airy mode's field is not evanescent anywhere in the cladding even though it is localised to the core-like region, and there will be a continuum of radiation modes of greater $\beta$ with power concentrated in the cladding. A dispersion plot of the fibre's modes shows anti-crossing behaviour, where the most "Airy-like" mode changes identity. These properties strongly suggest that the Airy mode is guided by a mechanism akin to photonic bandgap guidance, even though photonic bandgap ideas played no role in the design process.

The method we used to design the fibre can be extended to many other cases where a fibre is needed that guides a given desired field distribution as a guided mode. For example, for a near-field top-hat beam (not reported here) our method suggests how to approximate the field to give a realistic index distribution, which ends up being similar to previously-reported fibres that guide such a mode.

**Funding**. European Union OPTICON Research Infrastructure for Optical/IR Astronomy (EU-FP7 226604); UK Science and Technology Research Council (STFC-PRD) (ST/K00235X/1).

**Acknowledgment**. The authors thank R. Haynes and P. Kern for useful discussions.